# Spectrometer to Explore Isotopologues of Lunar Volatiles on Luna-27 Lander

Viacheslav Meshcherinov[a]*, Iskander Gazizov[a,b], Viktor Kazakov[a], Maxim Spiridonov[a], Yuri Lebedev[a], Imant Vinogradov[a], Mikhail Gerasimov[a]

[a] *Space Research Institute of the Russian Academy of Sciences (IKI RAS), 117997 Moscow, Russia*
[b] *Institute of Chemical Technologies and Analytics, TU Wien, 1060 Vienna, Austria*
*Corresponding author. Email: meshcherinov@phystech.edu

**Abstract**

The study of volatiles and the search for water are the primary objectives of the Luna-27 mission, which is planned to land on the south pole of the Moon in 2028. Here we present the tunable Diode Laser Spectrometer (DLS-L) that will be onboard the lander. The DLS-L will perform isotopic analysis of volatiles that are pyrolytically evolved from regolith. This article dives into the design of the spectrometer and the characterisation of isotopic signature retrieval. We look forward to expanding our knowledge of Lunar geochemistry by measuring D/H, $^{18}O/^{17}O/^{16}O$, $^{13}C/^{12}C$ ratios *in situ*, which could be the one-of-a-kind direct study of the lunar soil isotopy *without* sample contamination.

## 1. Introduction

*The Moon is a new continent awaiting its development*, as once said by Erik Galimov – researcher of the lunar regolith and originator of the Luna-25 mission [1]. No matter the world, every development starts with the exploration of the soil. We believe that the investigation of volatiles in regolith is one of the fundamental steps in understanding the Moon as a whole, as it tells the story of the lunar bowels, impact events, and interactions with the solar wind [2]. Moreover, geological exploration of the Moon could contribute to the reconstruction of the Earth's early history, when oceans, atmosphere, and complex organic compounds formed [1,3]. And with future lunar development missions, the study of volatiles is more of a practical interest [4,5]. In this article, we present a Diode Laser Spectrometer (DLS-L) that will provide information on the Moon's geological structure with *in situ* measurements of $H_2O$ and $CO_2$ isotopologues in the polar region of the Moon.

### 1.1. Lunar exploration brief

Preservation of volatile substances on the lunar surface is greatly complicated by unfavourable conditions, as the temperature ranges from -160ºC to +120ºC depending on the illumination. Wherein water could stay condensed on the lunar surface only lower than -166ºC, and the limit for $CO_2$ is even lower. However, as early as the 1960s, it was suggested that volatiles might be stored in the condensed state in the permanently shaded areas at the floor of polar craters – *'cold traps'* [6] where the temperature would not rise above -230ºC [7]. This assumption was confirmed only half a century later with the LCROSS experiment on the NASA LRO orbiter. During this experiment, the LEND neutron detector observed hydrogen content in the polar regions of the Moon, which corresponds to ice water with an estimated content of 0.5 to 4.0% [8].

Later, NASA's Moon Mineralogy Mapper on the ISRO Chandrayaan-1 orbiter constructed the first global quantitative maps of lunar surface water based on near-infrared reflectance data [9]. Water content was found to increase with latitude, approaching values of 500-750 ppm in polar regions. Detailed laboratory studies of lunar soil became possible after Soviet sample return missions Luna-16, -20, -24, and NASA Apollo missions.

There are three main possible sources of lunar volatiles: degassing of the lunar mantle, interaction of solar wind protons with surface rocks, and impact degassing of falling meteorites and comets [10]. The composition of the formed volatiles and their isotopologues is unique in each case:

A. The isotopic ratio of volatiles corresponds to the *lunar average.* In such cases, gases might escape through crustal cracks after degassing from hot magma in the lunar bowels. Laboratory studies of lunar regolith show δD ranging from +179 to +5420‰ with an average range of +187 to 340‰ [11,12,13].
B. The isotopic ratio of water corresponds to the source with a *solar isotopy of hydrogen and a lunar isotopy of oxygen*. In such cases, water molecules could be produced by the interaction of oxygen-containing components of the lunar regolith with high-energy protons of the solar wind in the absence of an atmospheric shield [14,15].
C. The isotopic ratio δD corresponds to the range *from deuterium-enriched solar wind to low thousands of ‰* [16,17]. In such cases, volatiles are released during the high-velocity impact of meteorites and comets on the lunar surface. The volatiles in the form of plumes spread over a large area of the lunar surface due to gas-dynamic expansion. The composition of the resulting volatiles is determined by equilibrium with heated silicate substances at high temperatures [18].

Even in early works on the analysis of samples returned to Earth, the authors noted the possibility of regolith contamination during transportation, storage, or laboratory analysis [19,20]. For example, detection of water in Luna-24 samples and its absence in previous missions could be explained either by contamination of the samples or by increased sensitivity of the instruments [20]. Studies of the carbon isotopy of regolith samples returned by the Luna-16 and Luna-20 missions showed the validity of concerns about contamination of the lunar soil [21]. In this way, the possibility of analysing lunar soil *in situ*, bypassing the return to Earth and possible contamination, would make it possible to validate previous geological measurements and help in future lunar exploration.

*1.2. Luna-27 mission*

With the new cycle of moon exploration, heirs of the resumed Soviet programs *Luna-25, -26, and -27* were announced. The DLS-L spectrometer presented in the article will be onboard the heavy lander Luna-27. The lander will perform contact studies of the Moon's volatile-rich South Pole, excavating the lunar regolith down to tens of centimetres. The declared active lifetime of the lander equipment on the lunar surface is at least one Earth year.

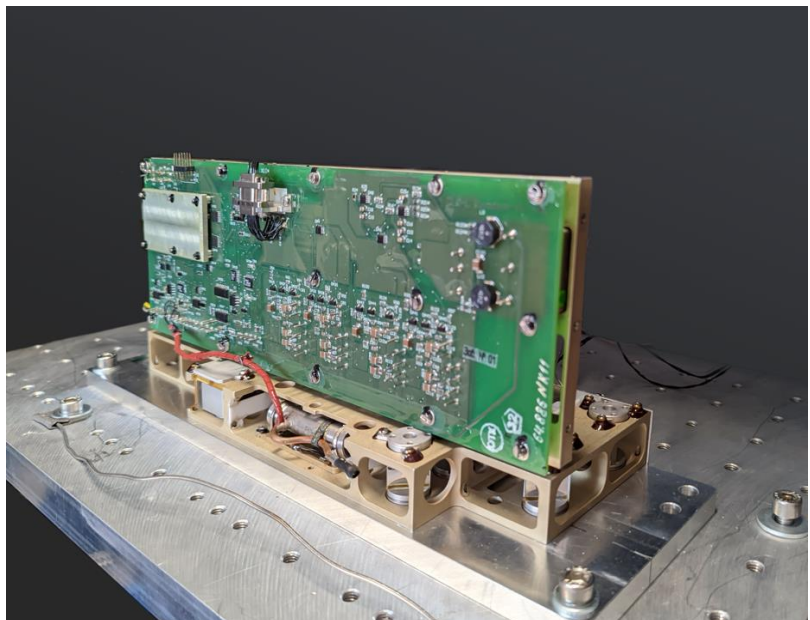

Fig. 1. Standalone DLS-L spectrometer during a calibration in a vacuum chamber.

The scientific equipment of the Luna-27 station includes the Gas Analysis Package (GAP), which is developed at the Space Research Institute of the Russian Academy of Sciences (IKI RAS) for a comprehensive analysis of lunar soil. The sample analysis lasts 40-90 minutes. It starts with collecting fine volatile-rich regolith from various depths by the compact multifunctional robotic arm of the Lunar Manipulator Complex (LMC), similar to the Luna-25 arm [22]. Next, the arm loads regolith samples into the TA-L instrument for direct thermal analysis by heating and pyrolytically producing volatiles. The TA-L instrument contains 8 high-temperature disposable pyrolytic cells that can heat up to 1000ºC to act on hydroxyl- and carbonates-containing substances. There is also a low-temperature reusable desorption device to target water and carbon dioxide adsorbed in the regolith. From TA-L, the released gases enter the Gas Chromatograph (GC-L) for measurements. Finally, volatile components enter the analytical cell of the Diode Laser Spectrometer (DLS-L) for real-time monitoring of the pyrolytic yield. But more importantly, DLS-L can perform isotopic analysis after gas preconcentration. The assembled DLS-L instrument is presented in Figure 1.

Preconcentration is a crucial step to improve the instrument's sensitivity. The volatiles are preconcentrated outside the DLS-L by two adsorption traps that contain a tube filled with Tenax adsorbent in one case, and the Carbosieve SIII adsorbent in the other. The adsorbent retains selected gases while cooling and releases them while heating. This system is similar to the one used in the Sample Analysis at Mars investigation of the Mars Science Laboratory Curiosity rover [23].

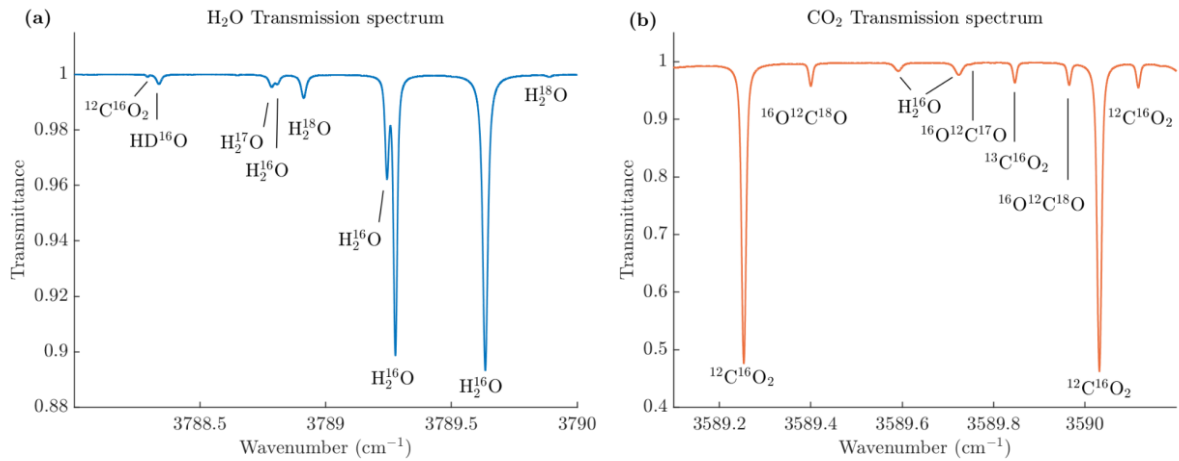

Fig. 2. Possible transmission spectra of lunar volatiles after regolith pyrolysis: (**a**) $H_2O$ at 3789 cm$^{-1}$ (2.639 μm) at a pressure of 15 mbar; (**b**) $CO_2$ at 3589.6 cm$^{-1}$ (2.786 μm) at a pressure of 8 mbar.

The DLS-L measures concentration dynamics of $H_2O$ and $CO_2$ during the pyrolytic yield, or performs isotope analysis of D/H, $^{18}O/^{17}O/^{16}O$, $^{13}C/^{12}C$. The instrument is based on the Tunable Diode Laser Absorption Spectroscopy (TDLAS) technique for its reliability and simplicity. This classic approach has proven itself well in space instrumentation [24,25]. The proposed spectral windows are presented in Figure 2 for a mixture of 50% $CO_2$ and 50% $H_2O$ with the possible lunar isotopologues abundances at a pressure of 30 mbar and a temperature of 30ºC [12,26,27]. In the first measurement scenario, we measure total abundance of $CO_2$ and $H_2O$ in a range of 2786.5 nm targeting the strongest absorption lines. In the second measurement scenario, we measure HDO, $H_2^{17}O$, and $H_2^{18}O$ isotopologues in the 2639 nm window; and $^{13}CO_2$ isotopologue in the 2786 nm window.

**Table 1**
Technical parameters of the DLS-L.

| Size, mm | Weight, g | Power consumption, W | Sensing volume, ml | Optical path, cm | Spectral range, cm$^{-1}$ | Targeted isotopologues |
|---|---|---|---|---|---|---|
| 258 × 88 × 115 | 650 | 5-7 | 1.34 | 19 | 3789 | HDO, $H_2^{17}O$, $H_2^{18}O$ |
| | | | | | 3589.6 | $^{13}CO_2$ |

As with any space mission, the defining parameters when creating the DLS-L spectrometer were *Mass*, *Power*, and *Size*. The available gas volume limit had the strongest impact on the final shape of the instrument. Based on the instrument's characteristics in Table 1, the DLS-L spectrometer could be classified as an ultra-compact sensor. However, the spectrometer had to provide reliable long-term measurements in a vacuum, under radiation, in the temperature range of ±50ºC, which is a complex task for non-laboratory equipment. Thereby, after three years of development, we present insights into instrument design, first cross-calibration, and our acquired experience for other teams.

## 2. Instrument Design

The DLS-L spectrometer is structurally embedded into the main GC-L instrument, sharing the gas, data, and power lines, as shown in Figure 3. At the time of writing, the qualification model of the instrument has not yet undergone Thermal-Vacuum, Vibration, and Shock tests. But the heir of this instrument which was installed on the Phobos-Grunt mission did successfully pass the preflight tests [28].

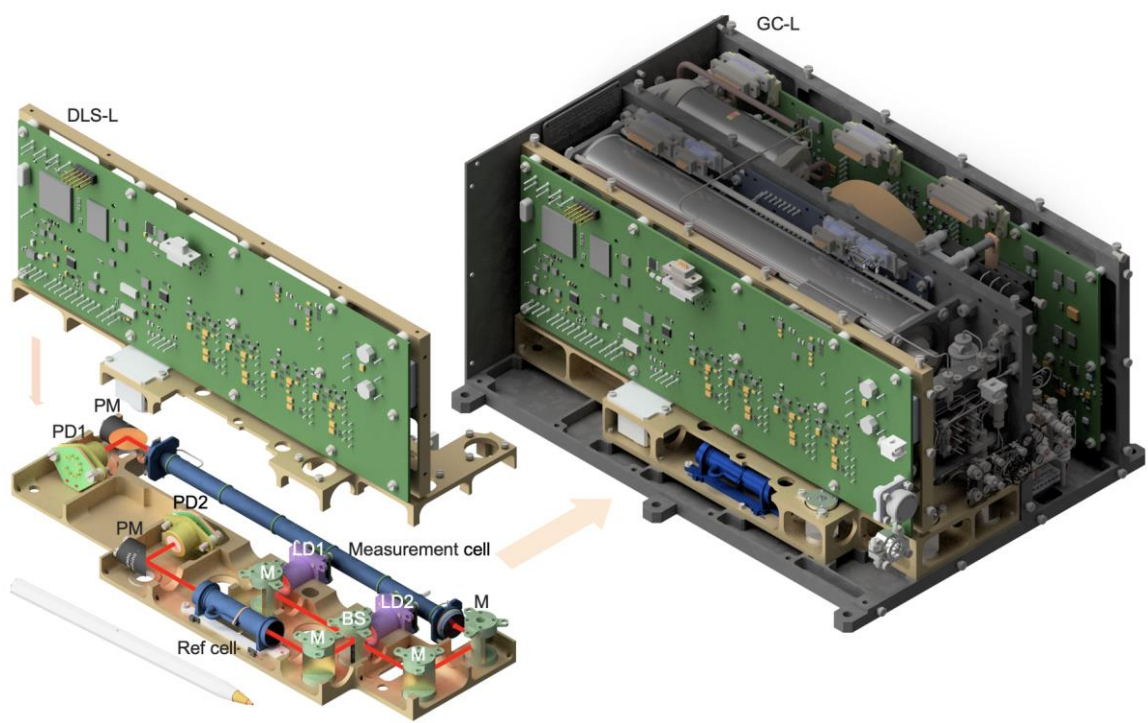

Fig. 3. Half-section view of DLS-L to present its optical bench, and diagram of DLS-L integrated into GC-L. M is a mirror; BS is a beamsplitter, PM is a parabolic mirror; Ref cell is a reference cell; PD is a photodiode; LD is a laser diode.

### 2.1. Optics

All the optical components are mounted on an anodized aluminium AMg6 body. The optical scheme of the instrument is presented in Figure 4. As a heart of the spectrometer, we employ two DFB diode lasers by Nanoplus in the TO5 package with an output power of 2-5 mW. Next, we collimate the beams with 4 mm focal length Thorlabs C036TME-D aspherical lenses. Collimated beams are combined with a 50/50 beamsplitter and guided to the stainless optical cell with $SiO_2$-protected aluminium-coated flat mirrors. This single-pass optical cell is made of a capillary tube 3mm in diameter and 190 mm in length with 8° tilted sapphire optical windows. The inner surface is passivated with a Sulfinert layer. Weak gas flow limits the optical cell volume, otherwise, the cell would act as a gas storage and disrupt normal circulation in the main GC-L instrument. After passing through the cell filled with 1.34 ml of analyte, the light is focused by Thorlabs MPD127127-90-M01 parabolic mirror on the J12TE3-66D-R01M InAs-photodetector from Teledyne Judson Technologies. There is an additional sealed single-pass reference cell with the same optical windows. This metal reference cell is filled with a mixture of 25% $H_2O$, 25% $CO_2$, and 50% $N_2$ at a pressure of 50 mbar at room temperature. The reference cell serves for verification purposes. All the described elements are permanently fixed with three-component epoxy K-400. All capillary tubes of the gas system and analytical cell are winded with insulated nichrome wire to heat up to 70ºC.

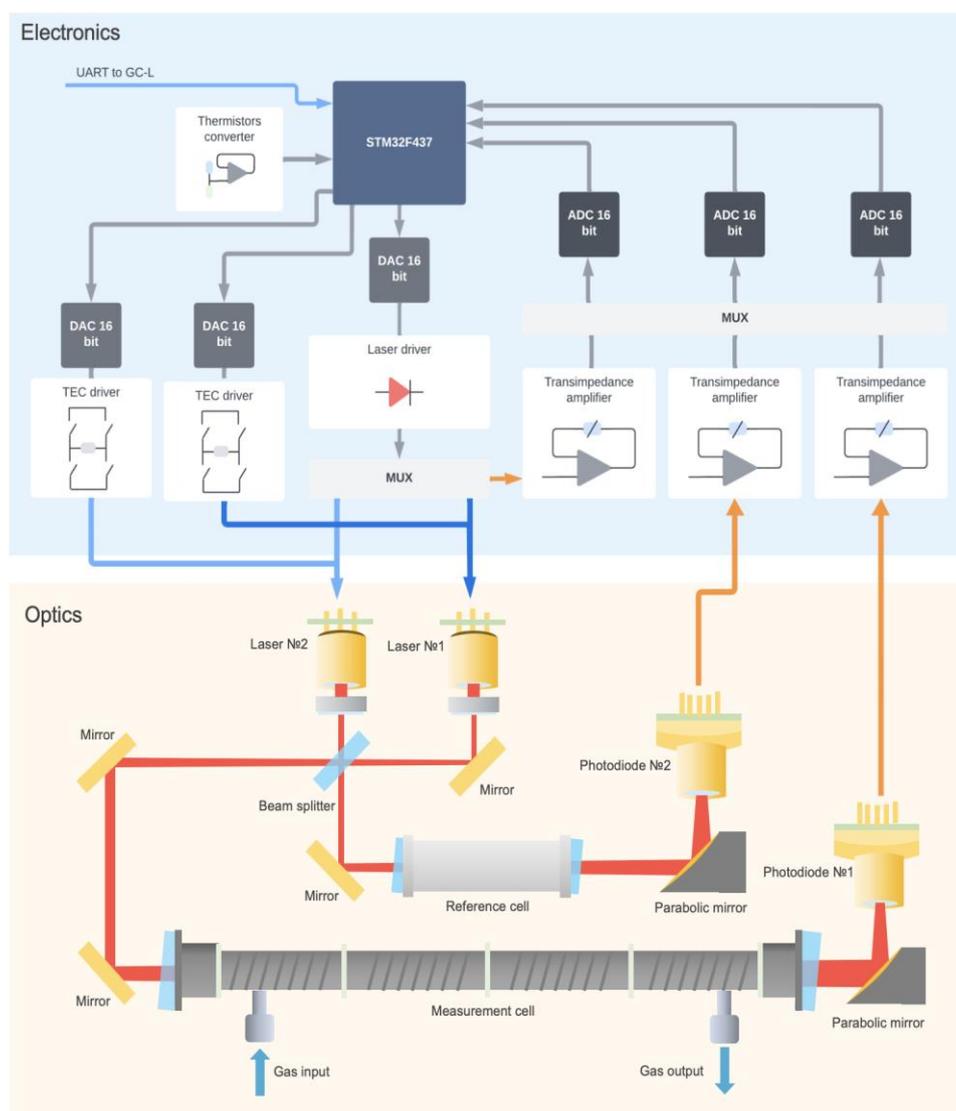

Fig. 4. Block diagram of electric circuits and principal optical diagram of the DLS-L instrument.

When designing the DLS-L we prioritised the overall robustness of the instrument, which led to the simplicity of the optical scheme and thus unavoidable fringing. We observe two types of fringes in measured spectra: high-frequency fringes from the inners of TO-5 laser package; and low-frequency fringes from optical components. To suppress low-frequency fringes we reduced back-reflections and implemented baseline correction in software. High-frequency fringes could be reduced mathematically or by averaging spectra in time.

*2.2. Electronics*

The STM32F437ZIT6 microcontroller (MCU) performs all data operations, communication, and control of peripheral devices. The MCU performance provides a continuous sampling rate of 128 kHz for tuning the laser current. A 16-bit DAC controls a laser current driver in the 0-200 mA range. The instrument has three transimpedance amplifiers (TIA) to convert the photocurrent of the analytical channel photodiode, reference channel photodiode, and monitor photodiode to the output signal voltage. The TIA gain is adjusted by digital code in the range of 5 kOhm to 2 MOhm. Next, three 16-bit ADCs digitise the TIA signals. The temperature of laser modules is controlled by a thermoelectric cooling (TEC) circuit with an analogue PID controller. The temperature is set with a 16-bit DAC in the range from 0 to 40°C. The temperature stability is 1 mK. The 12-bit ADCs integrated into the MCU digitise the board supply voltage and temperatures from thermistors, which are located on the reference cell, instrument frame, photodiodes, and laser module cases. The average power consumption of the device in the operating mode is less than 5.5 W. The General diagram of electronics is shown in Figure 4.

*2.3. Acquisition*

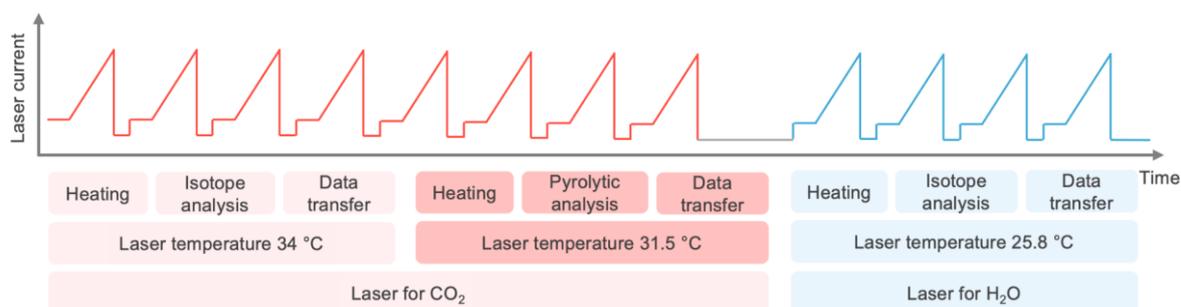

Fig. 5. Timeline option of a single acquisition cycle.

DLS-L measures spectra with two lasers in alternating mode, where each laser operates for 30 seconds, while the other one is not emitting. During the first 17 seconds the laser is modulated with a saw-tooth current shape and the inner TEC is settling on a specified temperature. This delay is required so the laser can get into stable condition, as we are also alternating laser temperatures to capture different spectral windows with the same laser. In the next 3 seconds, DLS-L records 32 spectra and averages them into one. Each spectrum is 10 ms long with 64 zero current points, 192 initial value points, and 1024 sawtooth tuning points. In the last 10 seconds, the 34.5 kB of measurements are transmitted through the 250 Kbit/s channel to the GC-L instrument. During an hour of operation, DLS-L measures 120 spectra with a volume of 4 MB. A single cycle of measurement is presented in Figure 5.

## 3. Results

We placed the DLS-L in a thermal vacuum chamber for all of the experiments in this article. For correct measurements, we had to reduce the influence of the water in the chamber and the instrument's gas system. This was done with a 3-day preparation routine for each experiment: we pumped out all of the volumes to a vacuum of ~$10^{-3}$ mbar according to the Atovac ACM200 pressure sensor, then purged the chamber with nitrogen, and repeated. We also measured water vapour separately from carbon dioxide to simplify the sample preparation procedure and avoid potential isotope exchange processes.

*3.1. Spectra processing*

The DLS-L instrument relies on the TDLAS technique for gas abundance retrieval. Data processing stages are as follows: instrument outputs internally averaged raw transmission spectrum once per minute, then we remove the baseline, determine the frequency axis for each spectrum, convert normalised transmission to absorption spectrum, and finally we average all spectra to reduce fast fringes. To determine the water and carbon dioxide concentration we fit the result with the gas mixture model.

Raw spectra measured by the DLS-L could be described by the Beer-Lambert relation:

$$I(\nu) = I_0(\nu)\ e^{-\sigma(\nu)\,N\,L}, \tag{1}$$

where $\sigma(\nu)$ is the cross-section (cm$^2$/molecules), $N$ is the concentration (molecules/cm$^3$), $L$ is optical path length (cm), $I(\nu)$ is received intensity, and $I_0(\nu)$ is the baseline, which is the optical signal in the absence of absorption. The baseline for each spectrum is retrieved either by directly fitting the non-absorbing spectral regions with a low-order polynomial [29] or with the orthogonal polynomials method [30]. The baseline contains not only the laser sweep shape but also low-frequency fringes, and ortho polynomials allow us to filter them both. After baseline removal, we convert the measured spectra to absorption coefficient $\alpha(\nu)$ (cm$^{-1}$) expressed in the following notation:

$$\alpha(\nu) = \sigma(\nu)\,N = \sigma(\nu)\,N_L\left(\frac{T_0}{T}\frac{p_{self}}{p_0}\right), \tag{2}$$

where $N_L$ is the Loschmidt constant (molecules/cm$^3$), $T$ is the temperature, $p_{self}$ is the partial pressure of a gas in the mixture, $T_0 = 273.15$ K, and $p_0 = 1$ atm.

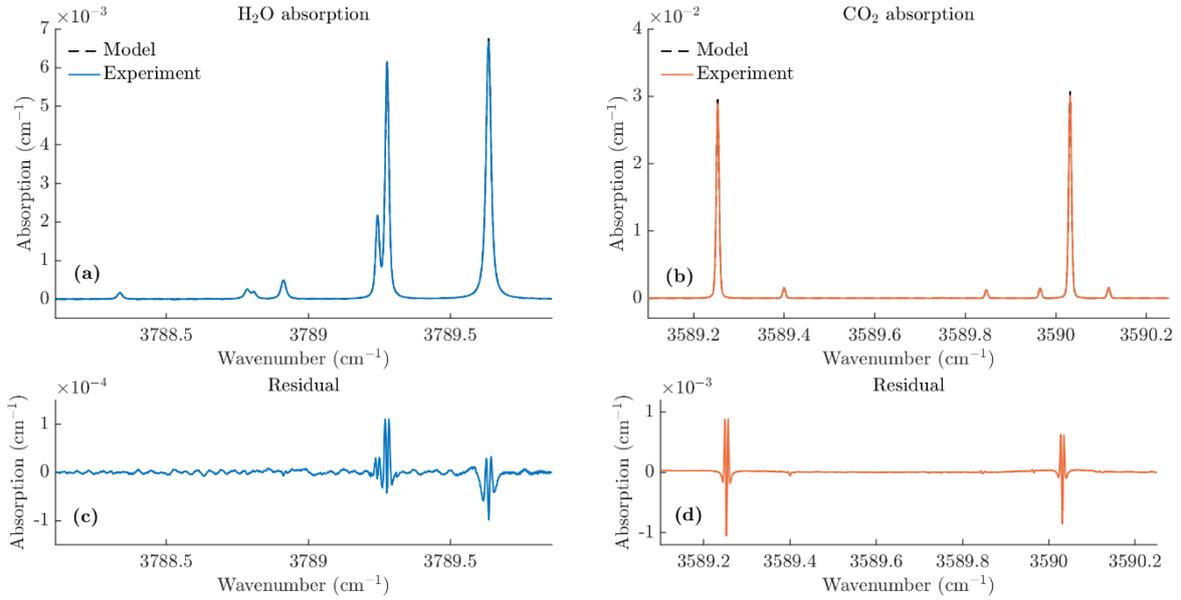

Fig. 6. Comparison of simulated and measured absorption coefficient and their residuals: (a) is H$_2$O absorption in the 2639 nm window, (b) is CO$_2$ in the 2786 nm window, (c) is H$_2$O residual, and (d) is CO$_2$ residual.

The next step is to link measured spectra to the physically accurate wavenumber axis. This is done by one-time calibration with a Fabry-Perot resonator, but further, we recalibrate each spectrum based on the positions of the absorption peaks to the values from the HITRAN database [31]. Finally, we fit measured absorption with the simulation to determine gas abundance. Line shape is based on the Voigt profile in the absence of parameters in the HITRAN-2020 database [31] for more precise Rautian [32], Galatry [33], or Hartmann-Tran [34] profiles. General concepts of absorption spectroscopy analysis have been repeatedly reviewed [35,36,37].

The resulting comparisons of simulated, measured, and residual absorption are presented in Figure 6. In the lab we measured H$_2$O in the range 3788-3789 cm$^{-1}$ at a pressure of 15 mbar and a temperature of 303 K; CO$_2$ in the range 3589.1-3589.3 cm$^{-1}$ at a pressure of 8 mbar and a temperature of 303.5 K.

*3.2. Isotopic ratios*

With the knowledge about concentrations for every isotopologue one could calculate the isotopic ratios as $R = \frac{N_{isotope}}{N_{main\ isotope}}$. More commonly the measured ratio of a sample is compared to the standard reference and their deviation is expressed by the isotopic signature δ:

$$\delta = \left(\frac{R_{sample}}{R_{standard}} - 1\right) \cdot 1000 \text{ ‰,} \quad (3)$$

where $R_{sample}$ is, for example, the measured $N_{HDO}/N_{H2O}$ ratio and $R_{standard}$ is the standard D/H ratio of Vienna Standard Mean Ocean Water (VSMOW).

Usually, an independent assessment is required to validate the quality of the isotopic signature measurements. There are standards of VSMOW and Vienna Pee Dee Belemnite (VPDB) for water and carbon dioxide. With the inaccessibility of these standards in our initial tests, we had to apply a different set of verification methods:

A. *Water vapour* signatures were verified with a sample of Perrier natural mineral water from the southeast of the French commune of Vergèze. This water source was selected for several reasons: 1) The water from this well has been characterized using Isotope Ratio Mass Spectrometers (IRMS) [38,39]; 2) The water is sourced consistently from the same well; 3) The composition of water remains stable due to the large volume of the underground source. Available publications are limited to isotopic signatures of HDO and $H_2^{18}O$, whereas the isotopic signature of $H_2^{17}O$ is not available due to the complexity of sample preparation. Thus, we define the value of $\delta^{17}O_{VSMOW} = 0.52 \times \delta^{18}O_{VSMOW}$ [40]. The sample was degassed before the experiment. Results are presented in Table 2.

B. *Carbon dioxide* signature was verified by testing a portion of gas from our cylinder on IRMS. Measurements were performed on a Delta XP by Thermo Fisher Scientific with high-temperature conversion elemental analyser TC/EA in Vernadsky Institute of Geochemistry and Analytical Chemistry (GEOKHI) of RAS. The comparison is presented in Table 2.

**Table 2**

Comparison of $H_2O$ isotopic signatures from DLS-L instrument with data from Perrier mineral water studies [38,39]. The $CO_2$ isotopic signatures from DLS-L are compared to the data from the mass spectrometer of GEOKHI RAS.

| Isotopic signature | Reviews [38,39] / GEOKHI RAS IRMS results | DLS-L results | DLS-L accuracy |
| --- | --- | --- | --- |
| $\delta_{VSMOW}^{18}O$ | -6.33 ± 0.02 ‰ | -13.23 ± 1.71 ‰ | 6.90 ‰ |
| $\delta_{VSMOW}D$ | -42.00 ± 0,20 ‰ | -54.34 ± 4.11 ‰ | 12.34 ‰ |
| $\delta_{VSMOW}^{17}O$ [40] | -3.28 ‰ | -4.20 ± 8.69 ‰ | 0.92 ‰ |
| $\delta_{VPDB}^{13}C$ | -47.55 ± 0.06 ‰ | -46.64 ± 1.12 ‰ | 0.91 ‰ |

We calculated the precision of the instrument for Table 2 with the Standard Error of the Mean (SEM). This statistical measure is suitable due to the time-dependent walk of the isotopic signature caused by the fringe movement, as discussed in the next chapter.

*3.3. Errors*

To estimate the precision for DLS-L in Table 2 we processed 30-40 absorption spectra measured in 1 hour and retrieved the isotopic signatures. The DLS-L instrument slightly underestimates isotopic signatures compared to the data known from the reviews [38,39]. The larger deviation of the measured $\delta_{VSMOW}D$ from the reference could be explained by a difference in the deposition rate of heavy water compared to normal water. This difference is amplified dramatically because of the huge surface area of our gas system.

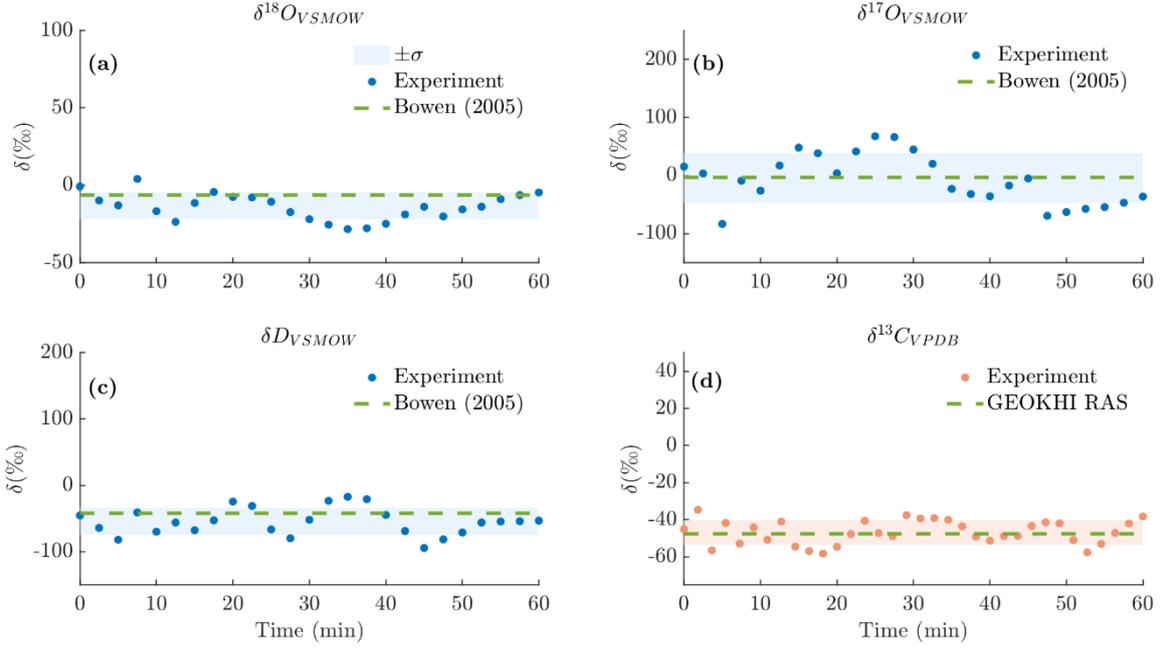

Fig. 7. Changes in measured isotopic signatures during the experiment: The dots show δ-values; green dashed lines show reference value; filled areas show 1$\sigma$ confidence interval. (**a**) is measured $\delta_{VSMOW}^{18}O$ compared to reviews [38,39]. (**b**) is measured $\delta_{VSMOW}^{17}O$ compared to reviews [38,39,40]. (**c**) is measured $\delta_{VSMOW}D$ compared to reviews [38,39]. (**d**) is measured $\delta_{VPDB}^{13}C$ compared to GEOKHI RAS results.

The accuracy of our retrieval is partly affected by the high-frequency fringes, which could result in an error of up to ~10‰ for $\delta_{VSMOW}^{18}O$, up to ~20‰ for $\delta_{VSMOW}^{17}O$, and up to ~30‰ for $\delta_{VSMOW}D$. However, due to the thermal cycling during the experiment, the fringe pattern shifts along the spectrum with an average speed of ~0.4 cm$^{-1}$ per hour. The period of the high-frequency fringes is ~0.04 cm$^{-1}$, which is close to the absorption widths. Thus, the fringe pattern shifts ~10 periods during an hour-long experiment, which strongly reduces its contribution to the inaccuracy of isotopic signatures. The behaviour of the measurements caused by fringe movement is presented in Figure 7.

**Table 3**
Instrument's SNR and LOD for the analysed isotopologues.

| Isotopologue | Signal-to-Noise Ratio | Limit of detection |
| --- | --- | --- |
| $H_2O$ | 800-2500 | 1,95×10$^{14}$ mol/cm$^3$ |
| $H_2^{18}O$ | 200 | 2,5×10$^{15}$ mol/cm$^3$ |
| $H_2^{17}O$ | 120 | 5,1×10$^{15}$ mol/cm$^3$ |
| HDO | 93 | 8,2×10$^{15}$ mol/cm$^3$ |
| $CO_2$ | 20 000 | 3,4×10$^{13}$ mol/cm$^3$ |
| $^{13}CO_2$ | 1250 | 8,2×10$^{14}$ mol/cm$^3$ |

We calculated the signal-to-noise ratio (SNR) and Limit of Detection (LOD) for every isotopologue in Table 3. $LOD_{iso}$ is defined as a minimal concentration of isotopologue $N_{iso}$, that could be distinguished from the noise of spectrum $\alpha_{noise\ spectrum}$:

$$LOD_{iso} = 3 \cdot min(N_{iso}) = 3\,\frac{std(\alpha_{noise\ spectrum})}{\sigma_{iso}}. \qquad (4)$$

## 4. Discussion

The aim of this paper was to present a design of a spectrometer to measure lunar volatiles in extremely harsh environments. After publishing this work, we are not finished with the development and testing of the instrument. The final version of the instrument will be assembled before the launch of the Luna-27 mission. In the flight-ready version of the instrument, we are introducing slight modifications. In software, the sample rate will be increased, and a wavelength stabilisation algorithm from our previous works will be implemented [25,41,42]. In instrument design, we plan to reduce fringes by employing higher-quality optical elements. In data processing, we might transition to complex line shapes like Hartmann-Tran. Since the HITRAN database is updated every 4 years, the additional line parameters for our spectral region might be available in the 2024 or 2028 version, given that IUPAC recommended the use of the Hartmann-Tran profile in its 2014 report [43]. Finally, we plan more tests of the instrument's performance and verification with certified standards before the flight in 2028.

With this paper, we propose a new instrument to perform *in situ* studies of lunar soil to avoid sample contamination, which might affect the returned samples during the years of storage and studies in laboratories. Despite strong limitations on the size, power, mass, and gas volume, the current performance does satisfy the requirements of the project. We conclude that DLS-L can measure HDO, $H_2^{18}O$, $H_2^{17}O$, and $^{13}CO_2$ isotopologues with the possibility of comparison to the results from Soviet missions Luna-16, Luna-20, Luna-24, and a series of NASA Apollo missions.

## 5. Acknowledgments

The authors gratefully acknowledge the team headed by Dr V.S. Sevastyanov from Vernadsky Institute of Geochemistry and Analytical Chemistry of RAS for supplying reference data for $CO_2$ sample isotopic ratios.

## 6. Funding

The work was supported by the IKI RAS grant "Planeta". The realisation of the device was supported by the State Corporation Roscosmos.